\documentclass[a4paper]{jpconf}
\usepackage{graphicx}
\usepackage{dcolumn}
\usepackage{bm}
\usepackage{amsmath}
\usepackage{color}

\begin{document}

\def\gsim{\mathop {\vtop {\ialign {##\crcr 
$\hfil \displaystyle {>}\hfil $\crcr \noalign {\kern1pt \nointerlineskip } 
$\,\sim$ \crcr \noalign {\kern1pt}}}}\limits}
\def\lsim{\mathop {\vtop {\ialign {##\crcr 
$\hfil \displaystyle {<}\hfil $\crcr \noalign {\kern1pt \nointerlineskip } 
$\,\,\sim$ \crcr \noalign {\kern1pt}}}}\limits}

\title{Magnetism and topological property in icosahedral quasicrystal}

\author{Shinji Watanabe}

\address{Department of Basic Sciences, Kyushu Institute of Technology, Kitakyushu, Fukuoka 804-8550, Japan}

\begin{abstract}
Quasicrystal (QC) has no periodicity but has a unique rotational symmetry forbidden in periodic crystals. Lack of microscopic theory of the crystalline electric field (CEF) in the QC and approximant crystal (AC) has prevented us from understanding the electric property, especially the magnetism. By developing the general formulation of the CEF in the rare-earth based QC and AC, we have analyzed the CEF in the QC Au-SM-Tb and AC (SM=Si, Ge, and Ga). The magnetic anisotropy arising from the CEF plays an important role in realizing unique magnetic states on the icosahedron (IC). By constructing the minimal model with the magnetic anisotropy, we have analyzed the ground-state properties of the IC, 1/1 AC, and QC. The hedgehog state is characterized by the topological charge of one and the whirling-moment state is characterized by the topological charge of three. The uniform arrangement of the ferrimagnetic state is stabilized in the QC with the ferromagnetic (FM) interaction, which is a candidate for the magnetic structure recently observed FM long-range order in the QC Au-Ga-Tb. The uniform arrangement of the hedgehog state is stabilized in the QC with the antiferromagnetic interaction, 
which suggests the possibility of
 the topological magnetic long-range order.
\end{abstract}

\section{Introduction}

The quasicrystal (QC) discovered in 1984~\cite{Shechtman} has no periodicity but has a unique rotation symmetry forbidden in periodic crystals. 
Although the understanding of the atomic structure of the QC has proceeded~\cite{Tsai,Takakura}, 
clarification of the electronic state and the physical property has still been fascinating and challenging problem since the QC lacks the translational invariance where the Bloch theorem can no longer be applied.
It has been a long-standing issue whether the magnetic long-range order is realized in the three-dimensional QC. 

In the approximant crystal (AC) which is the periodic crystal with the common local configuration of atoms to that of the QC, the magnetic long-range order was observed. In the 1/1 AC 
Cd$_6$R (R=Nd, Sm, Gd, Tb, Dy, Ho, Er, and Tm), the antiferromagnetic (AFM) order was observed~\cite{Tamura2010,Mori,Tamura2012,Das}. In the 1/1 AC Au-SM-R (SM=Si, Ge, Sn, and Al, and R=Gd, Tb, Dy, and Ho), the ferromagnetic (FM) order and the AFM order were observed~\cite{Hiroto2013,Hiroto2014,Sato2019,Hiroto}. 
In the 2/1 AC (Au, Cu)-(Al, In)-R (R=Gd, Tb), the FM order was observed~\cite{Inagaki}. In the 2/1 AC Ga-Pd-Tb, the AFM order was observed~\cite{So}. 
Experimental efforts to find the magnetic long-range order in the ACs until recently are well summarized in Ref.~\cite{Suzuki}. 

In the QC Cd$_6$R (R=Gd, Tb, Dy, Ho, Er, and Tm), spin-grass behavior was observed but magnetic long-range order was not observed~\cite{Goldman}. 
Recently, ferromagnetic (FM) long-range order has been discovered in QC Au-Ga-Gd and Au-Ga-Tb by Tamura et al~\cite{Tamura2021}. The magnetic susceptibility and specific heat measurements exhibit the transition temperature of 23 K and 16 K, respectively. In both systems, positive Curie-Weiss temperature was observed, which indicates the FM interaction working between 4f magnetic moments at the rare-earth sites.

\begin{figure}[t]
\begin{center}
\includegraphics[width=10cm]{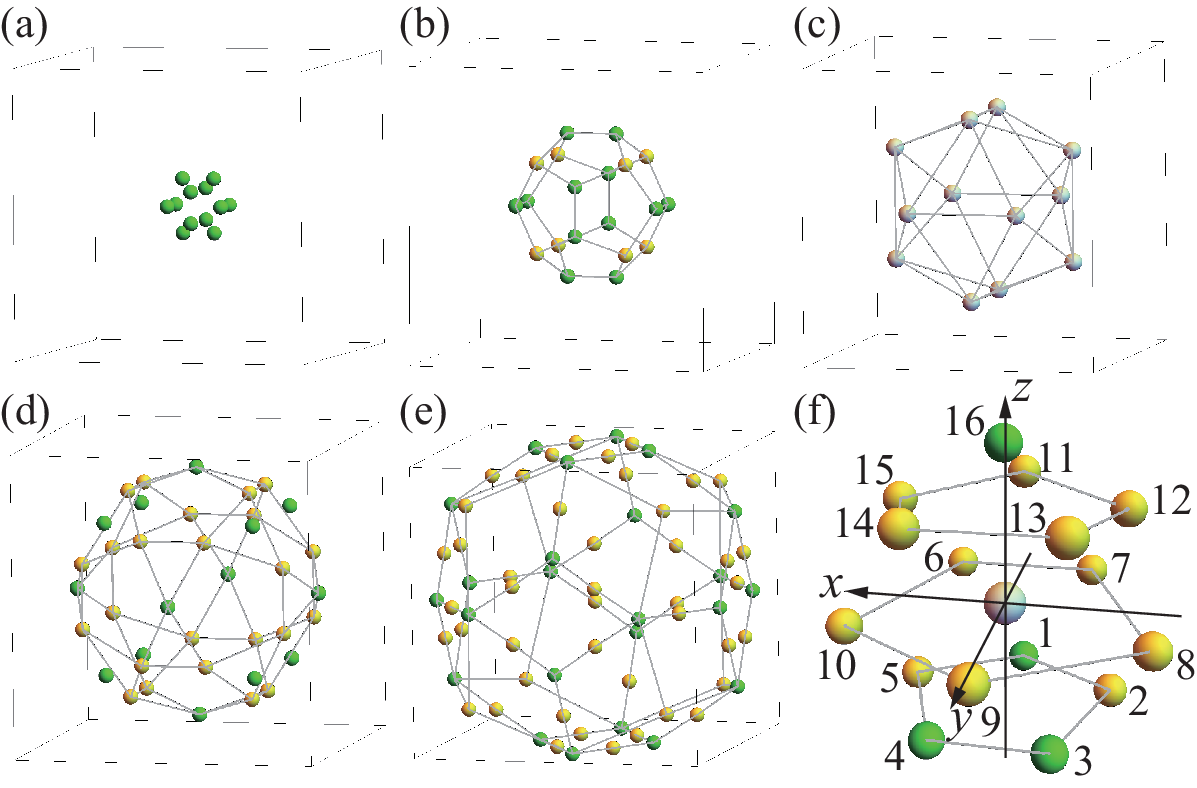}
\caption{\label{fig:Tsai}Tsai-type cluster consists of (a) cluster center, (b) dodecahedron, (c) icosahedron, (d) icosidodecahedron, and (e) defect rhombic triacontahedron with Tb (gray), Au (yellow), and Au/SM (green).
(f) Local configuration of atoms around the Tb site in the Tsai-type cluster. Surrounding Au and SM atoms are labeled with numbers. 
}
\end{center}
\end{figure}

The QCs as well as the ACs mentioned above, where the magnetic long-range order was observed, are composed of the atomic cluster named the Tsai-type cluster. 
The Tsai-type cluster consists of concentric shell structures shown in Figs.~\ref{fig:Tsai}(a)-\ref{fig:Tsai}(e). 
Here we show the atomic structure identified in the 1/1 AC Au$_{70}$Si$_{17}$Tb$_{13}$~\cite{Hiroto} as a typical case. 
The rare-earth atoms are located at 12 vertices of the icosahedron (IC) shown in Fig.~\ref{fig:Tsai}(c). 

So far, theoretical studies on magnetism in the QC as well as AC have been performed as the model calculations mostly by the spin model and also by the Hubbard model in a small cluster or low-dimensional systems~\cite{Okabe,Sorensen,Coffey,Jan,Axe,Wessel,Kons,Jan2007,Hucht,Thiem,Komura,Sugimoto,Koga2017,Koga2020,Koga2021,Koga2022,STS,Miyazaki}, where the effect of the crystalline electric field (CEF) crucial for the rare-earth system was not taken into account microscopically. 
It is noted that in Ref.~\cite{Sugimoto}, the magnetic anisotropy was introduced phenomenologically and the FM order in the 1/1 AC was shown to appear. 

\section{Crystalline electric field in rare-earth based QC and AC}

To clarify the magnetic anisotropy microscopically, understanding of the CEF is necessary. The CEF is important for 4f electron states in rare-earth based compounds. The CEF theory under crystalline point group compatible with periodicity is well established. However, lack of microscopic theory for CEF in rare-earth based QC and AC has prevented us from understanding the 4f electron states, especially, the magnetic properties. 

So far, experimental studies on the CEF were reported in the Tm- and Tb-based 1/1 ACs~\cite{Das,Jazbec,Hiroto} and the microscopic theory for the CEF in the QC and AC has been highly desired. 
Recently, the CEF Hamiltonian in the rare-earth based QC and AC has been formulated on the basis of the point charge model~\cite{WK2021}. 

The local configuration of the atoms around the rare-earth atom in the Tsai-type cluster is shown in Fig.~\ref{fig:Tsai}(f). 
Here, the atomic configuration in the 1/1 AC Au$_{70}$Si$_{17}$Tb$_{13}$ is shown as a typical case and 
the atoms at the cluster center are neglected since their existence probability is very low~\cite{Hiroto}. 
The vector passing through the Tb site from the center of the IC is set to be the $z$ axis, which is the pseudo-5 fold axis. 
The Tb site is set to be the origin. Then, the $y$ axis is set so as the $yz$ plane to be the mirror plane and the $x$ axis is set to be perpendicular to both the $z$ axis and the $x$ axis. 

In the Tsai-type cluster, the electrostatic interaction between the 4f electron at the position ${\bm r}$ with the charge $q$ and surrounding ligand ions at the position ${\bm R}_i$ with the charge $q_i$ in Fig.~\ref{fig:Tsai}(f) are expressed by
\begin{eqnarray}
H_{\rm CEF}=q\sum_{i}\frac{q_i}{|\bm{R}_i-\bm{r}|}. 
\label{eq:HCEF1}
\end{eqnarray}
Here, the ligand ions are labeled with numbers $i=1,\cdots,16$ as in Fig.~\ref{fig:Tsai}(f). 
Since the rare-earth based QC and AC are metallic crystals, the valences of the ligand ions are considered to be understood on the basis of the free-electron model whose nominal values are listed in the literature~\cite{Suzuki}. In reality, these are considered to be reduced by the screening effect of electrons. Hence, we set the valences of the screened ligand ions as $Z_{\rm Au}$ for the Au ion and $Z_{\rm SM}$ for the SM ion. 
Then, the charge of the Au ion and the SM ion are expressed as $q_i=Z_{\rm Au}|e|$ and $q_i=Z_{\rm SM}|e|$ respectively where $|e|$ is the elementary charge. 

By using the Wigner-Eckart theorem, eq.~(\ref{eq:HCEF1}) is expressed as 
\begin{eqnarray}
H_{\rm CEF}= 
\sum_{\ell=2,4,6}\left[B_{\ell}^{0}(c)O_{\ell}^{0}(c)+\sum_{\eta=c,s}\sum_{m=1}^{\ell}B_{\ell}^{m}(\eta)O_{\ell}^{m}(\eta)\right], 
\label{eq:HCEF2}
\end{eqnarray}
where $B_{\ell}^{m}$ is the coefficient and $O_{\ell}^{m}$ is the Stevens operator which is composed of the operators of the total angular momentum $\hat{J}_{+}=(\hat{J}_x+\hat{J}_y)/2$, $\hat{J}_{-}=(\hat{J}_x-\hat{J}_y)/(2i)$, and $\hat{J}_z$~\cite{WK2021}. 
The explicit forms of $B_{\ell}^{m}$ and $O_{\ell}^{m}$ are given in Ref.~\cite{WK2021}. 
In periodic crystals, due to the high symmetry of the crystalline point group, it has been well known that only a few terms for $\eta=c$ contribute to $H_{\rm CEF}$. 
However, in the rare-earth based QC and AC with (pseudo) 5-fold symmetry, all the terms with even $\ell$ for $2\le\ell\le 6$ appear in $H_{\rm CEF}$. 

By applying this formulation to the QC Au$_{51}$Al$_{34}$Yb$_{15}$, we analyzed the CEF considering the effect of Al/Au mixed sites~\cite{WK2021}. 
It was shown that the shape of the spherical part of the wavefunction of the CEF ground state $|\psi_0\rangle$, i.e., $\psi_0(\hat{\bm r})=\langle\hat{\bm r}|\psi_0\rangle$ changes depending on the valences of the screened ligand ions surrounding the Yb site~\cite{WK2021}. 

Next, to get insight into the CEF in the QC Au-Ga-Tb, we have analyzed the CEF at the Tb site in the Tsai-type cluster~\cite{WSR,WPNAS}. 
The Tb$^{3+}$ ion with $4f^8$ configuration is regarded to have the charge $q=6|e|$ in the hole picture since $4f^{14}$ configuration is the closed shell of the 4f electrons. The Hund's rule tells us that the ground multiplet is specified by the total angular momentum $J=6$. 
Then we have diagonalized eq.~(\ref{eq:HCEF2}) in the basis of $|J=6, J_z\rangle$ for $J_z=-6, -5, \cdots, 5 ,6$.

We treat the ratio of the valences of the screened ligand ions $\alpha=Z_{\rm SM}/Z_{\rm Au}$ as a parameter and 
set $Z_{\rm Au}=0.223$ as a typical value, which was identified in the 1/1 AC Au$_{70}$Si$_{17}$Tb$_{13}$ by the neutron measurement~\cite{Hiroto}. 
We have calculated the $\alpha$ dependence of the CEF energy and eigenstate by assuming that all the Au/SM mixed sites are occupied by SM~\cite{WPNAS}. 
We have also calculated the CEF energy and the eigenstate by considering the effect of the Au/SM mixed sites~\cite{WSR}. 

\begin{figure}[tb]
\begin{center}
\includegraphics[width=11cm]{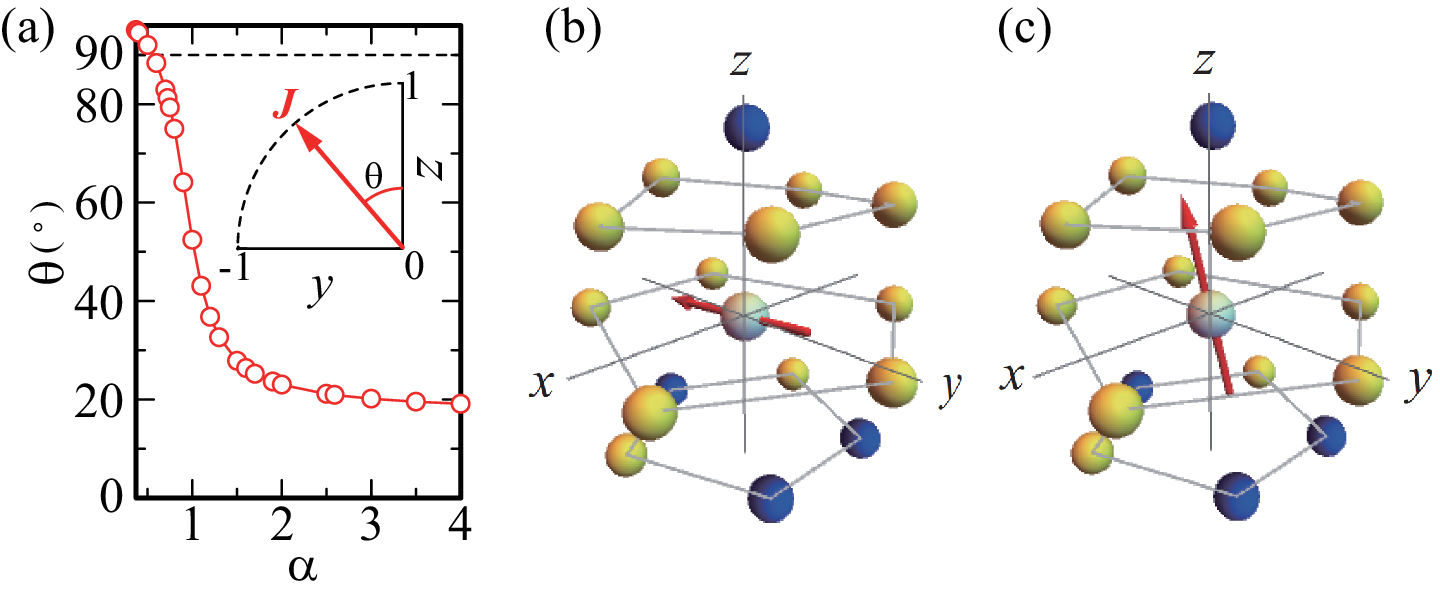}
\caption{\label{fig:CEF}
(a) The $\alpha$ dependence of the magnetic anisotropy. Inset illustrates the magnetic easy axis in the mirror $(yz)$ plane in the local coordinate shown in Fig.~\ref{fig:Tsai}(f). 
The magnetic easy axis with the anisotropy angle of (b) $\theta=95.14^{\circ}$ for $\alpha=0.365$ and of 
(c) $\theta=19.12^{\circ}$ for $\alpha=4.000$ in the local coordinate with Tb (gray), Au (yellow), and SM (blue). 
}
\end{center}
\end{figure}

To clarify the principal axis of the magnetic moment in the CEF ground state, we have calculated $3\times 3$ matrix $M_{\xi,\zeta}\equiv\langle\psi_0|\hat{J}_{\xi}\hat{J}_{\zeta}|\psi_0\rangle$ for $\xi, \zeta=x, y,$ and $z$~\cite{WSR,WPNAS}. 
By diagonalizing $M$, we have obtained the normalized eigenvector ${\bm J}$ for the largest eigenvalue, which gives the largest-moment direction. 
The result is shown in Fig.~\ref{fig:CEF}(a) in the case where the Au/SM mixed sites are occupied by Au~\cite{WPNAS}. 
For $\alpha\ge 0.365$, ${\bm J}$ is lying in the $yz$ plane in the local coordinate shown in Fig.~\ref{fig:CEF}(f), which is the mirror plane.
The magnetic easy-axis direction is defined by the angle $\theta$ of the ${\bm J}$ vector from the pseudo 5-fold axis i.e., the $z$ axis as shown in the inset of Fig.~\ref{fig:CEF}(a). 
At $\alpha=0.365$, the magnetic easy axis is nearly perpendicular to the pseudo 5-fold axis with $\theta=95.14^{\circ}$ as shown in Fig.~\ref{fig:CEF}(b). As $\alpha$ increases, the magnetic easy axis rotates to anticlockwise direction in the mirror plane (i.e., $yz$ plane) and approaches the pseudo 5-fold axis as shown in Fig.~\ref{fig:CEF}(c) where $\theta=19.12^{\circ}$ at $\alpha=4.000$. 
It is noted that the magnetic easy axis lying in the mirror plane was indeed observed in the 1/1 AC Au$_{72}$Al$_{14}$Tb$_{14}$ with $\theta=86^{\circ}$~\cite{Sato2019} and Au$_{70}$Si$_{17}$Tb$_{13}$ with $\theta=80^{\circ}$~\cite{Hiroto}. 

It is noted that for $\alpha<0.365$, we have obtained ${\bm J}=(1,0,0)$ which is perpendicular to the mirror plane i.e., the $yz$ plane in Fig.~\ref{fig:CEF}(f)~\cite{WPNAS}. This magnetic anisotropy has not been identified experimentally. Experimental observation of this anisotropy is left for interesting future study.

\section{Minimal model for magnetism}

The development of the microscopic theory of the CEF has made it possible to reveal the magnetic anisotropy as shown in the previous section. Then, we construct the minimal model for 4f magnetism. The rare-earth based ACs and QCs are metallic compounds. The distance between the rare-earth atoms are larger than 5 angstrom so that RKKY interaction mediated by conduction electrons is considered to work between the 4f magnetic moments. The effect of Fermi surface or pseudo Fermi surface such as the nesting and number of conduction electrons can also contribute to the magnetism. 
Keeping these effects in mind, here, as a first step of analysis, we employ the effective model taking into account the effect of the CEF anisotropy. 
\begin{eqnarray}
H=-\sum_{\langle i,j \rangle}J_{ij}\hat{\bm J}_i\cdot\hat{\bm J}_j,
\label{eq:H}
\end{eqnarray}
where $\hat{\bm J}_i$ is the unit-vector operator parallel or anti-parallel to magnetic easy axis and $J_{ij}$ is taken as nearest-neighbor (N.N.) interaction $J_1$ and next nearest neighbor (N.N.N.) interaction $J_2$. We consider the case that magnetic easy axis is lying in the mirror plane. 

It should be noted that the model (\ref{eq:H}) was introduced by T. J. Sato to analyze the 1/1 AC Au$_{72}$Al$_{14}$Tb$_{14}$~\cite{Sato2019} where the model (\ref{eq:H}) applied to the icosahedron (IC) was succeeded in explaining main features of the measured temperature dependence of the magnetic susceptibility as well as the magnetic-field dependence of the magnetization. 

We have applied the model (\ref{eq:H}) to the IC, 1/1 AC, and the QC both with the FM interaction~\cite{WPNAS} and AFM interaction~\cite{WSR}. We have shown that various magnetic states appear in the ground-state phase diagram in the plane of $J_2/J_1$ and the magnetic anisotropy $\theta$. This implies that the magnetic anisotropy arising from the CEF plays an essential role in realizing unique magnetic structure on the IC. Notable is that our result for the FM interaction explains the FM long-range order of the ferrimagnetic state [see Fig.~\ref{fig:QC_ferri}(b)] observed in the 1/1 AC Au$_{70}$Si$_{17}$Tb$_{13}$~\cite{Hiroto} and the AFM long-range order of the whirling and anti-whirling moment states [see Figs.~\ref{fig:HH}(e) and \ref{fig:HH}(f)] observed in the 1/1 AC Au$_{72}$Al$_{14}$Tb$_{14}$~\cite{Sato2019}. This implies that the minimal model (\ref{eq:H}) captures the essential feature of real materials, which is indeed regarded as the effective model.

\section{Topological magnetic texture on IC}

We have shown that various magnetic states appear in the ground state of the IC~\cite{WSR,WPNAS}. Among them, we have found topological magnetic textures which are characterized by the non-trivial topological invariant, i.e., the topological charge. Here we discuss the qualitative feature of the topological magnetic texture on the IC in \S4.1 and topological transition in the 1/1 AC in \S4.2.

\subsection{Topological charge on IC}

\begin{figure}[tb]
\begin{center}
\includegraphics[width=16cm]{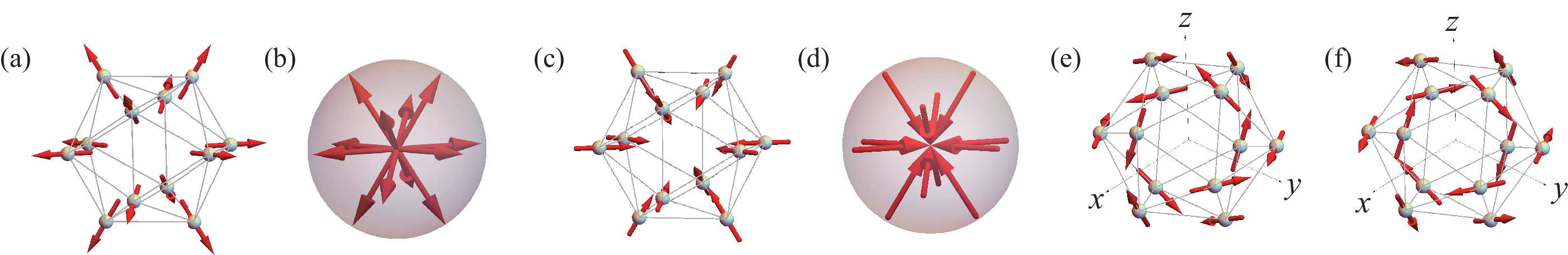}
\caption{\label{fig:HH}(a) The hedgehog state on the IC. (b) The 12 magnetic moments in the hedgehog state are gathered at the center of the unit sphere. 
(c) The anti-hedgehog state on the IC. 
(d) The 12 magnetic moments in the anti-hedgehog state are gathered at the center of the unit sphere. 
(e) The whirling moment state and (f) the anti-whirling moment state on the IC seen from the (111) direction. 
}
\end{center}
\end{figure}

By defining the topological charge on the IC $n\equiv\Omega/(4\pi)$ which is the ratio of the solid angle subtended by 12 magnetic moments on the IC and surface area of the unit sphere, we have revealed that the hedgehog state on the IC shown in Fig.~\ref{fig:HH}(a) is characterized by the topological charge of one $n=1$~\cite{WSR,WPNAS}. 
The topological charge indicates the number of covering whole sphere by the magnetic moments on the IC. 
As clearly seen in Fig.~\ref{fig:HH}(b), it is understandable that the topological charge of the hedgehog state is one. 
The anti-hedgehog state is the state with all magnetic moments being inverted from that of the hedgehog state as shown in Fig.~\ref{fig:HH}(c). 
The anti-hedgehog state is characterized by topological charge of $n=-1$ [see Fig.~\ref{fig:HH}(d)], which has the same energy as that of the hedgehog state. The hedgehog state is the source of emergent field, which is regarded as monopole with the charge $n=+1$ and anti-hedgehog state is the sink of emergent field, which is regarded as anti-monopole with the charge $n=-1$. 
The whirling-moment state is shown in Fig.~\ref{fig:HH}(e). 
We have revealed that the whirling-moment state is characterized by the unusually large topological charge of three $n=+3$~\cite{WSR,WPNAS}. 
The anti-whirling-moment state is the state with all magnetic moments being inverted from that of the whirling-moment state as shown in Fig.~\ref{fig:HH}(f). The anti-whirling-moment state is characterized by the topological charge $n=-3$. 
The whirling-moment state and anti-whirling-moment state are also energetically degenerate. 
In these states, the total magnetic moment for the IC is zero, ${\bm J}_{\rm tot}=\sum_{i=1}^{12}\langle\hat{\bm J}_i\rangle={\bm 0}$. 

Although the model (\ref{eq:H}) is introduced to describe the Ising-type interaction with uniaxial anisotropy arising from the CEF as the most relevant interaction, in real systems it is noted that there also exists the transverse interaction for the magnetic moments in addition to Eq. (\ref{eq:H}). Hence, the topological charge $n$ is regarded as an invariance under the continuous transformation of the magnetic moment in real systems. 

\subsection{Field-induced metamagnetic and topological transition in 1/1 AC}

In the 1/1 AC, magnetic field is applied to the hedgehog-anti-hedgehog AFM order~\cite{WPNAS}.
Then, field-induced metamagnetic transition occurs simultaneously with the discontinuous charge in the topological charge $|n|=1\to n=0$. 
Interestingly, topological Hall effect is expected to appear after the metamagnetic transition~\cite{WPNAS}. 
The field-induced metamagnetic transition also appears in the AFM order of whirling-anti-whirling-moment state in the 1/1 AC~\cite{WPNAS}. 
At the metamagnetic transition, the topological charge changes as $|n|=3\to 0$. The topological Hall effect is expected to appear after the metamagnetic transition. The metamagnetic transition was actually observed in the 1/1 AC Au$_{72}$Al$_{14}$Tb$_{14}$~\cite{Sato2019}. Hence, it is interesting if our theoretical prediction about the topological Hall effect is examined experimentally in the future study.

\section{Magnetism in icosahedral QC}

\begin{figure}[tb]
\begin{center}
\includegraphics[width=16cm]{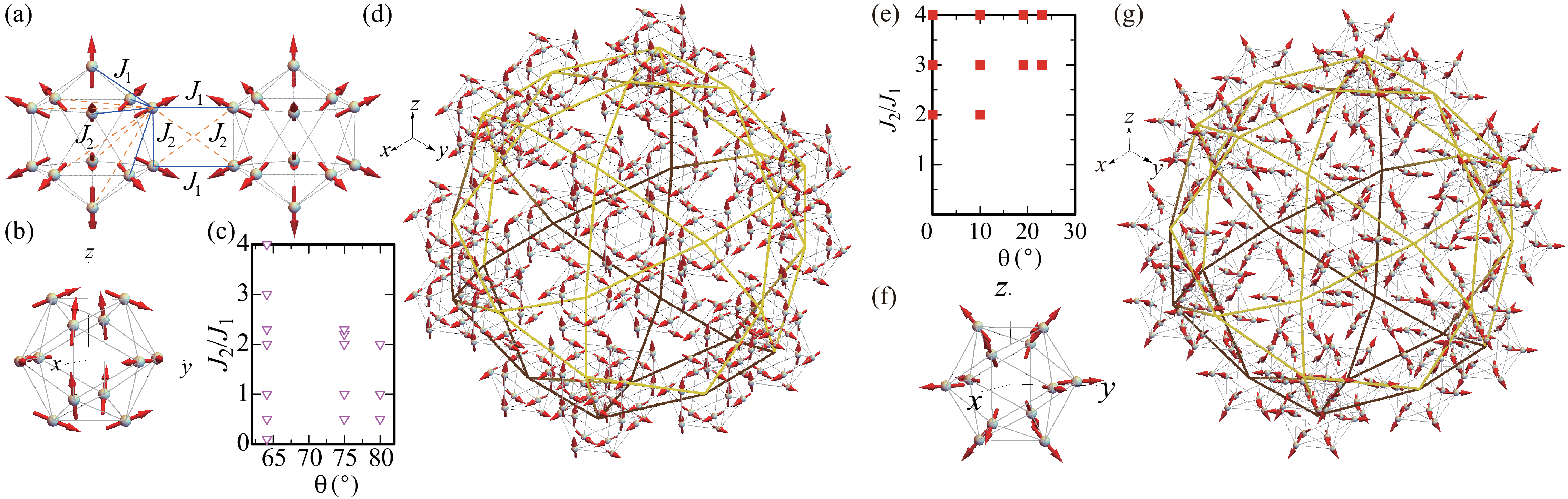}
\caption{\label{fig:QC_ferri}(a) The N.N. interaction $J_1$ (solid line) and the N.N.N. interaction $J_2$ (dashed line) in the QC.
(b) Ferrimagnetic state for $\theta=80^{\circ}$ on the IC. 
(c) The uniform ferrimagnetic state (open inverted triangle) realized in the ground-state phase diagram for $J_1>0$. 
(d) Uniform arrangement of the ferrimagnetic state on the IC for $\theta=64^{\circ}$. 
Green (brown) lines at the front (back) side connect the vertices of the icosidodecahedron. 
(e) The uniform hedgehog state realized in the ground-state phase diagram of the QC (red square) for $J_1<0$. 
(f) The hedgehog state for $\theta=10^{\circ}$ on the IC.
(g) Uniform arrangement of the hedgehog state on the IC for $\theta=10^{\circ}$. 
Green (brown) lines at the front (back) side connect the vertices of the icosidodecahedron. 
}
\end{center}
\end{figure}

To get insight into magnetism in the QC, 
we have applied the minimal model (\ref{eq:H}) to the Cd$_{5.7}$Yb-type QC~\cite{Takakura}. 
As a first step of analysis, we consider 30 ICs located at the vertices of icosidodecahedron. 
Since the lattice structure of the QC Au$_{65}$Ga$_{20}$Tb$_{15}$ was not solved experimentally, here we employ the icosidodecahedron observed in the 1/1 AC Au$_{70}$Si$_{17}$Tb$_{13}$~\cite{Hiroto}. Namely, we set the IC shown in Fig.~\ref{fig:Tsai}(c) at each vertex of the $\tau^3$-times enlarged icosidodecahedron shown in Fig.~\ref{fig:Tsai}(d), where $\tau$ is the golden mean $\tau=(1+\sqrt{5})/2$. 
Then, we consider the model (\ref{eq:H}) with the N.N. interaction $J_1$ and the N.N.N. interaction $J_2$ not only for intra IC but also for inter IC [see Fig.~\ref{fig:QC_ferri}(a)] for each bond connecting the N.N. vertices of the icosidodecahedron. The number of the neighboring pairs of the ICs is 60 and the total number of the Tb sites is $N=12\times 30=360$. 
We have evaluated the inter IC energy $\langle\hat{\bm J}_i\cdot\hat{\bm J}_j\rangle$ for the 360 Tb sites under the open boundary condition. 

First, we have calculated the FM interaction case for $J_1>0$ and $J_2>0$ in the model (\ref{eq:H})~\cite{WPNAS}. 
Then we have found that the ferrimagnetic state shown in Fig.~\ref{fig:QC_ferri}(b) is energetically favorable since the inner product of the N.N.N. magnetic moments are all parallel $\langle\hat{\bm J}_i\cdot\hat{\bm J}_j\rangle=1$ for all the inter IC bonds [see Fig.~\ref{fig:QC_ferri}(a)] of the icosidodecahedron. 
This implies that at least in the large $J_2/J_1$ region where the ferrimagnetic state is realized in the ground-state phase diagram of the IC, uniform arrangement of the ferrimagnetic state is energetically favorable. 
Indeed we have confirmed that the uniform arrangement of the ferrimagnetic state is stabilized as the ground state in the region of Fig.~\ref{fig:QC_ferri}(c), as shown in Fig.~\ref{fig:QC_ferri}(d)~\cite{WPNAS}. 
As seen in Fig.~\ref{fig:QC_ferri}(b), 
the total magnetic moment of the IC ${\bm J}_{\rm tot}=\sum_{i=1}^{12}\langle\hat{\bm J}_i\rangle$ is finite for this state, which is directed to the $(111)$ direction. 
Hence, uniform distribution of the ferrimagnetic state shown in Fig.~\ref{fig:QC_ferri}(d) gives rise to the finite magnetization along the $(111)$ direction. 
This state is considered to be a candidate for the magnetic structure of the FM long-range order recently observed in the QC Au-Ga-Tb~\cite{Tamura2021}. It is interesting to compare this result with future neutron measurement to identify the magnetic structure of the FM state in the QC Au-Ga-Tb. 
We note that the topological charge of the ferrimagnetic state on the IC [see Fig.~\ref{fig:QC_ferri}(b)] is zero.

We have also calculated the AFM interaction $J_1<0$ and $J_2<0$ in the model (\ref{eq:H})~\cite{WSR}. Then we have found that the uniform arrangement of the hedgehog state is energetically favorable since the inner product of the N.N.N. magnetic moments are all antiparallel $\langle\hat{\bm J}_i\cdot\hat{\bm J}_j\rangle=-1$ for all the inter IC bonds [see Fig.~\ref{fig:QC_ferri}(a)]. 
This implies that at least in the large $J_2/J_1$ region where the hedgehog state is realized in the ground-state phase diagram of the IC, uniform arrangement of the hedgehog state is energetically favorable. 
Indeed we have confirmed that the uniform arrangement of the hedgehog state is stabilized as the ground state phase diagram shown in Fig.~\ref{fig:QC_ferri}(e). Interestingly, the hedgehog state appears in the broad region ranging from $\theta=0^{\circ}$ to finite $\theta$. 
The hedgehog state on the IC for $\theta=10^{\circ}$ is shown in Fig.~\ref{fig:QC_ferri}(f) and the uniform arrangement in the QC is shown in Fig.~\ref{fig:QC_ferri}(g)~\cite{WSR}. 


Finally,  we remark the CEF at each Tb site inside the Tsai-type cluster in the Cd$_{5.7}$Yb-type QC. 
The CEF is considered to be dominated by the N.N. ligand ions surrounding the Tb site, as shown in Fig.~\ref{fig:Tsai}(f). From the crystal viewpoint, the environment of each Tb site is different site by site. However, from the physical viewpoint, the ligand ions distanced from each Tb site is considered to give minor contribution to the CEF since such far ions are screened by other electrons, which is irrelevant to the electrostatic interaction. Moreover, the wavefunctions of the valence electron at such far ions have no overlap with the 4f wavefunction at the Tb site, which is also irrelevant to the hybridization picture of the CEF. Although very minor contribution from  further  neighbor ligand ions may not be completely negligible, the present analysis  taking into account the N.N. ligand ions is considered to capture the most relevant feature of the magnetic easy axis of each Tb site. 

\section{Summary}

We have developed the theory of the CEF on the basis of the point charge model and have analyzed the CEF in the QC Au-SM-Tb and AC. 
We have shown that the magnetic anisotropy arising from the CEF plays an essential role in realizing unique magnetic structure on the IC. 
By constructing the minimal model with the uniaxial magnetic anisotropy, we have analyzed the ground state properties of the IC and the 1/1 AC, and the QC. We have shown that the hedgehog state is characterized by the topological charge of one and the whirling moment state is characterized by the topological charge of three. In the QC, we have found that the uniform arrangement of the ferrimagnetic state is stabilized for the FM interaction, which is a candidate for the magnetic structure of the FM long-range order recently observed in the QC Au$_{65}$Ga$_{20}$Tb$_{15}$. 
We have also found that the uniform arrangement of the hedgehog state is stabilized in the QC for the AFM interaction, 
which suggests the possibility of
the topological magnetic long-range order in the QC. 

\section*{Acknowledgements}

The author thanks collaboration with M. Kawamoto and S. Mie. 
The author acknowledges R. Tamura, S. Suzuki, and T. Ishimasa for informing me of their experimental results as well as valuable discussions. 
This work was supported by JSPS KAKENHI Grant Numbers JP18K03542, JP19H00648, JP22H0459, and JP22H01170.

\end{document}